\documentclass{elsart}
\usepackage{amssymb}
\newcommand{\Tr}{\mathop{\rm Tr}\nolimits}
\journal{Physics Letters A}
\begin{document}
\begin{frontmatter}

\title{Bogolyubov approximation for diagonal model of an interacting Bose gas}
\author[Chile]{M. Corgini },
\ead{mcorgini@yahoo.com}
\author[Russia]{D.P. Sankovich\corauthref{cor}}
\corauth[cor]{Corresponding author.}
\ead{sankovch@mi.ras.ru}

\address[Chile]{Departamento de
Matem\'aticas, Universidad de La Serena,\\ Cisternas 1200, La
Serena, Chile}
\address[Russia]{V.A. Steklov Mathematical Institute,
Gubkin str. 8, 119991 Moscow, Russia}

\begin{abstract}
We study, using the Bogolyubov approximation, the thermodynamic
behaviour of a superstable Bose system whose energy operator in
the second-quantized form contains a nonlinear expression in
the occupation numbers operators. We prove that for all values of the chemical potential satisfying $\mu > \lambda(0)$, where $\lambda (0)\leq 0$
is the lowest energy value, the system undergoes Bose--Einstein condensation.
\end{abstract}

\end{frontmatter}

1. Exactly solvable models of strongly interacting bosons could be
helpful in understanding the nature of Bose--Einstein condensation
(BEC) and superfluidity in interacting Bose gases. In this paper we
study the thermodynamic behavior of some theoretically relevant
system of Bose particles. The considered model has rather simplified
character. Researches related to similar models are connected with
the attempt to consider the effect of excitation-excitation coupling in
Bogolyubov theory.

The aim of this paper is to consider a model of Bose gas with diagonal interaction which shows an independent on temperature BEC.
This kind of BEC was theoretically discovered by Bru and Zagrebnov for some specific Bose system with diagonal interactions ~\cite{bru2,bru3}. Here we study a nonlinear modification of the Bru--Zagrebnov model.
Furthermore we want to clear up the influence of high-order perturbations in the non-ideal Bose gas interaction on the magnitude of the Bose condensate fraction.

The plan of the paper is as follows. Firstly, we describe the basic
mathematical notions associated to this kind of systems. Secondly,
we discuss the thermodynamic behavior of a superstable Bose system
whose Hamiltonian contains a nonlinear term in the number operators.
The nonlinearity can be physically understood as the simultaneous
creation of $k$-single one mode particles after the disappearance of
an equivalent amount of single particles associated to the same
mode. For $k=2$ a variation of this kind of models has been
extensively studied in refs.~\cite{bru2,bru3,bru4,ber} by using
different mathematical techniques. (In contrast to the Bru--Zagrebnov model ~\cite{bru3}, the Hamiltonian we consider contains a specific nonlinear mean-field term (see also ~\cite{van})). In our case use will be made of
the so-called Bogolyubov approximation ~\cite{bog}. In our proof a
significative role plays the fact that the involved Hamiltonian is
written in terms of the occupation numbers operators. It enables us
to give a simplified proof of thermodynamic equivalence of the limit
grand-canonical pressures corresponding to the energy operator of
the system  and its respective approximative Hamiltonian.

We have to point out that although this modified approach leads to a
straightforward proof of BEC based on the commutativity of the
involved operators, it does not apply for proving spontaneous
breakdown of symmetry.

2. Let $\Delta$ be the operator Laplacian. The one-particle free
Hamiltonian corresponds to the operator $ D_{\Lambda_l } =
-\Delta/2$ defined on a dense subset of the Hilbert space
${\mathcal{H}}^l = L^2 (\Lambda_l)$, being $\Lambda_l = [ -l/2,
l/2]^d \subset\mathbb R^d $ a cubic box of boundary $\partial
\Lambda_l$ and volume $ V_l= l^d $. We assume periodic boundary
conditions under which $ D_{\Lambda_l }$ becomes a self-adjoint
operator. In this case, as in ref.~\cite{bru4,lbz}, we shall
consider a negative isolated lowest value of energy. This can be
done by adding a term to the original Hamiltonian which shifts the
energy level of the zero mode of the kinetic energy operator
downward by a negative amount, creating in this way  a spectral gap.

We shall study  model of interacting Bose particles whose
Hamiltonian is given as
\begin{equation}
\label{ham1}
H_l =  H^0_l + \frac{\gamma_0}{V^{k-1}_l}(a^{\dag}_0)^k
a^k_0 +\frac{\gamma_1}{V^{k-1}_l}\displaystyle\sum_{p\in {\Lambda
}^*_l\backslash\{0\}} (a^{\dag}_p)^ka^k_p + V_l
g\left(\frac{{\hat{N}}^{'}}{V_l}\right),
\end{equation}
where integer $k\ge 2$ and $ g(x) $ is a suitable real valued
function satisfying general conditions ensuring the superstability
of the model. In particular, it suffices to assume that $g(x)$  is a
continuously differentiable function on $[0,\infty)$ satisfying $g(0)=0$
and $g^{'}(x) \to\infty $ as $x\to\infty$ ~\cite{dav}. The sum in (\ref{ham1}) runs over the set $\Lambda^*_l
= \{ p\in {\mathbb R}^d: p_{\alpha} = 2\pi n_{\alpha}/l, n_{\alpha}
= 0,\pm 1,\pm 2, \dots ,\alpha =1,2,\dots,d \} $, $ a^{\dag}_p, a_p$
are the Bose operators of creation and annihilation of particles
defined on the Fock space ${\mathcal{F}}_{\mathrm B}$ satisfying the
usual Bose commutation rules: $[a_p, a^{\dag}_q] = a_pa^{\dag}_q -
a^{\dag}_qa_p = \delta_{p,q}, $ being $\delta_{p,q}$ the Kronecker's
delta. Denote by $ n_p = a^{\dag}_pa_p$ the occupation-number
operator in the mode $p\in\Lambda^*_l$. Let $\lambda_l(p)= p^2/2$
for $ p\neq 0$ and $\lambda_l(0)\equiv \lambda (0) \leq 0 $. In this
case $ H^0_l = \sum_{p\in {\Lambda }^*_l} \lambda_l(p)n_p $,
$\gamma_0 ,\gamma_1 > 0.$ $N = \sum_{p\in {\Lambda }^*_l} n_p$ is
the total number operator and $ N^{'} = \sum_{p\in {\Lambda
}^*_l\backslash \{ 0 \} }n_p$ is the number operator with exclusion
of $n_0.$ Note that the boson Fock space ${\mathcal{F}}_{\mathrm B}$
is isomorphic to the tensor product $ \otimes_{p\in \Lambda^*_l}
{\mathcal{F}}^{\mathrm B}_p $ where ${\mathcal{F}}^{\mathrm B}_p $
is the boson Fock space constructed on the one-dimensional Hilbert
space $ {\mathcal{H}}_p = \{ \gamma e^{ipx}\}.$

Let
\[
p_l(\beta, \mu) = \frac{1}{\beta V_l}\ln \Tr_{\otimes_{p\in
{\Lambda}^*_l} {\mathcal{F}}^{\mathrm B}_p}
 \exp [ -\beta( H_l - \mu N)]
\]
be the grand-canonical pressure corresponding to $H_l$, where $\mu$
represents the chemical potential and $ \beta = \theta^{-1}$ is the
inverse temperature.

If $ H_l(\mu)\equiv  H_l - \mu N$, the equilibrium Gibbs state
$\langle - \rangle_{H_l(\mu)}$ is defined as
\[
\langle A \rangle_{H_l(\mu)} \equiv \left[\Tr_{\otimes_{p\in
{\Lambda}^*_l} {\mathcal{F}}^{\mathrm B}_p} \exp\left(-\beta H_l
(\mu)\right)\right]^{-1}\Tr_{\otimes_{p\in {\Lambda}^*_l}
{\mathcal{F}}^{\mathrm B}_p} A \exp\left(-\beta H_l (\mu)\right),
\]
for any operator $A$ acting on the symmetric bosonic Fock space. The
density of particles at infinite volume is defined as
\[
\rho (\mu)=
\lim_{V_l\to\infty}\rho_{l}(\mu)\equiv\lim_{V_l\to\infty}\left\langle
\frac{N}{V_l}\right\rangle_{H_l(\mu)}.
\]
The density of particles of energy $\lambda (0) $, corresponding to
the mode $p=0$, is given as
\[
\rho_{0}(\beta, \mu) = \lim_{V_l\to\infty}\rho_{0,l}(\beta,\mu)
\equiv\lim_{V_l\to\infty}\left\langle
\frac{n_0}{V_l}\right\rangle_{H_l(\mu)}.
\]

In this paper we want to prove the occurrence of macroscopic
occupation of the ground state, i.e. we attempt to find values of
$\mu$ and $\beta$ for which the inequality $ \rho_{0}(\beta, \mu) >
0$ holds. Indeed, we prove for the studied model that macroscopic
occupation independent on temperature takes place, i.e. the
density of the condensate not depends on $\beta$.

3. In the framework of the Bogolyubov approximation we must replace
the operators $ a^{\dag}_0 $ and $a_0 $ in any operator $A$
expressed in the normal form in the Hamiltonian $H_l (\mu)$ with the
complex numbers $\sqrt{V_l}\bar{c}$ and $\sqrt{V_l}c $. We get the
following approximative Hamiltonian

\begin{eqnarray}
\label{ham2} H^{\mathrm {appr}}_l (c,\mu) =
\sum_{p\in\Lambda^*_l\backslash\{0\}} \left[(\lambda_l(p)-\mu )n_p +
\frac{\gamma_1}{V^{k-1}_l}(a^{\dag}_p)^k a^k_p\right]+ V_l
g\left(\frac{N^{'}}{V_l}\right) \nonumber\\
+ (\lambda (0)-\mu ) V_l |c|^2 + \gamma_0 V_l |c|^{2k},\;\;c\in
\mathbb C.
\end{eqnarray}
Noting that
$
(a^{\dag}_{p})^{k}(a_{p})^{k} = n_p(n_p-1 )\cdots (n_p - k + 1 )
$
for all $p\in \Lambda^*_l $, it is not hard to see that both
Hamiltonians (\ref{ham1}) and (\ref{ham2}) are diagonal with respect
to the occupation-number operators.

In this case we have the following result~\cite{gin}.

\begin{thm}\label{thm1}
\[
\Tr_{\otimes_{p\in \Lambda^*_l\backslash \{0\}}
{\mathcal{F}}^{\mathrm B}_p} e^{-\beta H^{\mathrm {appr}}_l (c,\mu)
}\leq \Tr_{\otimes_{p\in \Lambda^*_l} {\mathcal{F}}^{\mathrm B}_p}
e^{-\beta H_l (\mu)  }.
\]
\end{thm}

4. Let $f_{V_l} (x)$ be the real valued function defined as
\[
f_{V_l} (x) = (\mu - \lambda (0))x - \gamma_0 x
\left(x-\frac{1}{V_l}\right) \cdots\left( x-\frac{k-1}{V_l}\right).
\]
We define the functions
\[
f^{\mathrm L} (x) = (\mu - \lambda (0))x - \gamma_0 x^k
\]
and
\[
f^{\mathrm U}_{V_l} (x) = (\mu - \lambda (0))x - \gamma_0 \left(x-
\frac{k-1}{V_l}\right)^k.
\]
For $x\ge (k-1)/V_l$ the functions $ f^{\mathrm L} (x) $ and
$f^{\mathrm U}_{V_l} (x) $ are concave with unique maxima attained
at $ x^{\mathrm L}$ and $ x^{\mathrm U}_{V_l}$, respectively.
Evidently,
\[
\lim_{V_l\to \infty} x^{\mathrm U}_{V_l} = x^{\mathrm L}
=\left[\frac{(\mu -\lambda (0))_+}{\gamma_0 k
}\right]^{\frac{1}{k-1}},
\]
where $(a)_+=\max(0,a)$.

\begin{lem}\label{lem1}
For $V_l$ sufficiently large such that $x^{\mathrm L} > (k-1)/V_l,$
the following inequalities
\[
f^{\mathrm L} (x^{\mathrm L}) \leq\sup_{[\frac{k-1}{V_l},\infty)
}f_{V_l}(x) \leq f^{\mathrm U}_{V_l}( x^{\mathrm U}_{V_l}),
\]
\[
\sup_{[0, \frac{k-1}{V_l}) }f_{V_l}(x) \leq (\mu -\lambda
(0))_+\frac{k-1}{V_l} +\gamma_0\frac{(2(k-1))!}{(k-2)!V^k_l}
\]
hold.
\end{lem}

\begin{pf}
For $V_l$ sufficiently large and $ x \in
\left[(k-1)/V_l,\infty\right) $ we have the obvious inequalities $
f^{\mathrm L} (x) \leq f_{V_l} (x) \leq f^{\mathrm U}_{V_l} (x)$,
which for $x^{\mathrm L} > (k-1)/V_l$ imply the first statement of
the proposition. On the other hand,
\[
f_{V_l}(x) \leq  (\mu - \lambda (0))_+x + \gamma_0 x
\left(x+\frac{1}{V_l}\right) \cdots\left(x+\frac{k-1}{V_l}\right)
\]
for $x\in [0,(k-1)/V_l)$. Then,
\[
\sup_{[0,  \frac{k-1}{V_l}  ) }f_{V_l}(x) \leq (\mu - \lambda
(0))_+\frac{k-1}{V_l}+\gamma_0 \frac{k-1}{V_l}\frac{k}{V_l}\cdots
\frac{2(k-1)}{V_l}.
\]
\end{pf}
\qed
\begin{lem}\label{lem2}
The following inequality takes place:
\[
\lim_{V_l\to \infty} \displaystyle\sup_{[0,\infty) }f_{V_l}(x) \leq
f^{\mathrm L} (x^{\mathrm L}) =\left[\frac{(\mu-\lambda
(0))_+}{k\gamma_0}\right]^{\frac{1}{k-1}}(\mu-\lambda (0))_+
\frac{k-1}{k}.
\]
\end{lem}

\begin{pf}
\[
|f^{\mathrm U}_{V_l}(x^{\mathrm U}_{V_l}) - f^{\mathrm L}
(x^{\mathrm L})|= f^{\mathrm U}_{V_l}\left( x^{\mathrm L} +
\frac{k-1}{V_l}\right) - f^{\mathrm L} (x^{\mathrm L}) =
(\mu-\lambda (0))_+\frac{k-1}{V_l},
\]
i.e. $\lim_{V_l\to \infty} f^{\mathrm U}_{V_l}(x^{\mathrm U}_{V_l})
= f^{\mathrm L} (x^{\mathrm L})$. Then, using this result and the
first inequality in lemma \ref{lem1}, we get
\[
\lim_{V_l\to \infty}\sup_{[\frac{k-1}{V_l},\infty) }f_{V_l}(x) =
f^{\mathrm L} (x^{\mathrm L}) =\left[\frac{(\mu-\lambda
(0))_+}{k\gamma_0}\right]^{\frac{1}{k-1}}(\mu-\lambda (0))_+
\frac{k-1}{k}.
\]
The second inequality of lemma \ref{lem1} yields
\[
\lim_{V_l\to \infty} \sup_{[0, \frac{k-1}{V_l}) }f_{V_l}(x)\leq 0.
\]
From these results, noting that
\[
\sup_{[0, \infty) }f_{V_l}(x) \leq \max\{\sup_{[0, \frac{k-1}{V_l} )
}f_{V_l}(x),\sup_{[\frac{k-1}{V_l},\infty) }f_{V_l}(x)\},
\]
and passing to the limit $V_l \to \infty$, the proof of the lemma
follows.
\end{pf}
\qed

Let $\hat{\rho}_0 =n_0/V_l$. $\hat{\rho}_0$ is a densely defined on
${\mathcal{F}}_0^{\mathrm B}$ positive self-adjoint operator with
spectrum $\sigma(\hat{\rho}_0 ) \subset [0,\infty)$ and spectral
family $ \{ E_{\hat{\rho}_0}(x): x\in \sigma(\hat{\rho}_0 )\}$
~\cite{reed}. Let $h(x): [0,\infty) \rightarrow\mathbb R$ be a
continuous function such that $\sup_{x\in [0,\infty)} h(x)$ exists.
Assume also that $ h(x)\exp(\beta V_l f_{V_l})$ is a bounded and
continuous function in $[0,\infty)$. Let $A, B$ be self-adjoint
bounded operators in the Hilbert space $\mathcal{U}$ endowed with
inner product denoted as $ (\cdot,\cdot)$. Then we say that $A$ is
smaller than $B$ and write $ A \leq B $ if $(\phi, A \phi)\leq
(\phi, B \phi)$ for every $\phi \in \mathcal{U}$.

\begin{lem}\label{lem3}
\[
\langle  h( \hat{\rho}_0 ) \rangle_{H_l(\mu)} \leq \sup_{x\in
[0,\infty)} h(x).
\]
\end{lem}
\begin{pf}
The spectral theorem for self-adjoint operators~\cite{reed} enables
us to write
\[
h( \hat{\rho}_0 )\exp(\beta V_l f_{V_l} (\hat{\rho}_0 )) =
\int_{\sigma(\hat{\rho}_0 )} h(x) \exp(\beta V_l f_{V_l}(x))\d
E_{\hat{\rho}_0}(x).
\]
Although $\hat{\rho}_0 $ is an unbounded operator, the above
representation is valid for any $\phi \in {\mathcal{F}}^{\mathrm
B}_0$ since $ h(x)\exp(\beta V_l f_{V_l} (x ))$ is a bounded
continuous function. This implies that,
\[
h( \hat{\rho}_0 )\exp(\beta V_l f_{V_l} (\hat{\rho}_0 )) \leq
\sup_{x\in [0,\infty)} h(x) \int_{\sigma(\hat{\rho}_0 )}\exp(\beta
V_l f_{V_l}(x))\d E_{\hat{\rho}_0}(x),
\]
in the sense of operators for all $\phi \in {\mathcal{F}}^{\mathrm
B}_0$ .
Noting that
\[ \exp(-\beta \hat{H}_l (\mu))= \exp(\beta V_l
f_{V_l} (\hat{\rho}_0 ))\exp(-\beta \hat{H}^{'}_l (\mu))
\] and since
$\exp(\beta V_l f_{V_l} (\hat{\rho}_0 ))$, $\exp(-\beta \hat{H}^{'}_l
(\mu)) $ are positively definite operators commuting with each
other, it is not hard to see that
\begin{eqnarray*}
\lefteqn{h(\hat{\rho}_0 )\exp(\beta V_l f_{V_l} (\hat{\rho}_0 ))\exp(-\beta
\hat{H}^{'}_l (\mu))}\\
&\leq \sup_{x\in [0,\infty)} h(x)\exp(\beta V_l
f_{V_l} (\hat{\rho}_0 ))\exp(-\beta \hat{H}^{'}_l (\mu)).
\end{eqnarray*}
Then, taking into account that $\hat{H}_l (\mu)$ defines a
superstable system and since ${\mathcal{F}}_{\mathrm B}$ is
isomorphic $ \otimes_{p\in \Lambda^*_l} {\mathcal{F}}^{\mathrm B}_p
$ we get
\begin{eqnarray*}
\lefteqn{\Tr_{{\mathcal{F}}_{\mathrm B}}h( \hat{\rho}_0 )\exp(\beta V_l f_{V_l}
(\hat{\rho}_0 ))\exp(-\beta \hat{H}^{'}_l (\mu))}\\
&\leq \sup_{x\in
[0,\infty)} h(x)\Tr_{{\mathcal{F}}_{\mathrm B}}\exp(\beta V_l f_{V_l}
(\hat{\rho}_0 ))\exp(-\beta \hat{H}^{'}_l (\mu)).
\end{eqnarray*}
Hence, we conclude that
\[
\langle h( \hat{\rho}_0 ) \rangle_{\hat{H}(\mu)} \leq \sup_{x\in
[0,\infty)} h(x).
\]
\end{pf}
\qed
\begin{thm}
\label{thm2} The systems with Hamiltonians $H_l(\mu)$ and $
H^{\mathrm {appr}}_l (c,\mu)$ are thermodynamically equivalent in
the Wentzel sense ~\cite{wen}, i.e.
\[
\lim_{V_l\to\infty} \sup_{| c |: c\in {\mathbb C}}  p^{\mathrm
{appr}}_l (\beta,c,\mu) = p(\beta,\mu),
\]
where
\[
p^{\mathrm {appr}}_l (\beta,c, \mu) \equiv \frac{1}{\beta V_l}\ln
\Tr_{\otimes_{p\in \Lambda^*\backslash\{0\}} {\mathcal{F}}^{\mathrm
B}_p} \exp [ -\beta( H^{\mathrm {appr}}_l(c,\mu))]
\]
and
$p(\beta,\mu) \equiv \lim_{V_l\to\infty} p_l (\beta,\mu).$
\end{thm}
\begin{pf}
Note that this is an equivalence of the Wentzel type but considering
infimum for the real variable $\left|c\right|$ instead  for $ c\in
{\mathbb C}$. Applying the Bogolyubov's convexity inequalities  for
pressures and theorem \ref{thm1} we obtain that
\[
0\leq p_l(\beta,\mu) -  p^{\mathrm {appr}}_l (\beta,c, \mu) \leq
\frac{1}{V_l}\langle  H^{\mathrm {appr}}_l (c,\mu) - H_l(\mu)
\rangle_{H_l(\mu)},
\]
where
\begin{eqnarray*}
\frac{1}{V_l}\langle  H^{\mathrm {appr}}_l (c,\mu) - H_l(\mu)
\rangle_{H_l(\mu)} =  (\lambda (0)-\mu ) |c|^2 + \gamma_0 |c|^{2k}\\
+ \left\langle(\mu-\lambda (0))\hat{\rho}_0 - \gamma_0\hat{\rho}_0
\cdots\left(\hat{\rho}_0 -
\frac{k-1}{V_l}\right)\right\rangle_{H_l(\mu)}.
\end{eqnarray*}
Hence, we obtain
\begin{equation}
\label{ineq} 0\leq p_l(\beta,\mu) - p^{\mathrm {appr}}_l (\beta,c,
\mu) \leq (\lambda (0)-\mu ) |c|^2 + \gamma_0 |c|^{2k} + \langle
f_{V_l} (\hat{\rho}_0 )\rangle_{H_l(\mu)}.
\end{equation}
We have to distinguish two cases.

(i) Case $\mu > \lambda (0) $. In this situation, $\inf_{| c |: c\in
{\mathbb C}} \left[(\lambda (0)-\mu ) | c |^2 + \gamma_0 | c
|^{2k}\right] $ is obtained for $|c_0|$ given by
\[
|c_0 |^2 = \left(\frac{\mu-\lambda
(0)}{k\gamma_0}\right)^{\frac{1}{k-1}}.
\]
This leads to
\begin{eqnarray*}
\inf_{| c |: c\in {\mathbb C}} \left[(\lambda (0)-\mu )
\left|c\right|^2 + \gamma_0 \left|c\right|^{2k}\right] = -
\left(\frac{\mu-\lambda
(0)}{k\gamma_0}\right)^{\frac{1}{k-1}}(\mu-\lambda (0))
\frac{k-1}{k}\\=-f^{\mathrm L} (x^{\mathrm L}).
\end{eqnarray*}
From the above result, using lemma \ref{lem3} with $h(x)=f_{V_l}(x)$
and considering the infimum for $c\in {\mathbb C}$ in the right-hand
side of inequality (\ref{ineq}), it follows that
\begin{eqnarray*}
0\leq p_l(\beta,\mu) - \sup_{| c |: c\in {\mathbb C}} p^{\mathrm
{appr}}_l (\beta, c, \mu)  \leq  - \left(\frac{\mu-\lambda
(0)}{k\gamma_0}\right)^{\frac{1}{k-1}}(\mu-\lambda (0))
\frac{k-1}{k}\\+\sup_{[0, \infty) }f_{V_l}(x).
\end{eqnarray*}
Therefore, passing to the limit $V_l\to\infty$ and applying lemma
\ref{lem2}, we get
\[
p(\beta,\mu)- \lim_{V_l\to\infty} \sup_{| c |: c\in {\mathbb C}}
p^{\mathrm {appr}}_l (\beta, c,\mu) = 0.
\]

(ii)Case $\mu \leq \lambda (0) $. In this case $\inf_{|c|: c\in
{\mathbb C}} \left[(\lambda (0)-\mu ) \left|c\right|^2 +  \gamma_0
\left|c\right|^{2k}\right]$ is attained at $|c_0| = 0$. Moreover,
since $(a^{\dag}_0)^ka^k_0$ is a positive operator, for $\mu \leq
\lambda (0)$ we obtain
\begin{eqnarray*}
\left\langle(\mu-\lambda (0))\hat{\rho}_0 -
\gamma_0\hat{\rho}_0\cdots \left(\hat{\rho}_0 -
\frac{k-1}{V_l}\right)\right\rangle_{H_l(\mu)}\\
=\left\langle (\mu - \lambda (0))\hat{\rho}_0
-\gamma_0\frac{(a^{\dag}_0)^ka^k_0}{V^k_l}\right\rangle_{H_l(\mu)}\leq
0.
\end{eqnarray*}
Then, from inequality (\ref{ineq}) we get,
\[
p_l(\beta,\mu) = \sup_{\left|c\right|: c\in {\mathbb C} }p^{\mathrm
{appr}}_l (\beta, c, \mu)
\]
for any finite $V_l$. Therefore, in both cases, passing to the limit
$V_l\to\infty$, we get
\[
p(\beta,\mu) = \lim_{V_l\to\infty} \sup_{|c|: c\in {\mathbb C}}
p^{\mathrm {appr}}_l (\beta, c,\mu).
\]
The proof is complete.
\end{pf}
\qed

5. The following theorem holds.
\begin{thm}
\label{thm3} For all $\mu>\lambda(0)$ the system (\ref{ham1}) displays independent on
temperature BEC and the amount of condensate is given by
\[
\rho_0(\mu) = \left(\frac{\mu -\lambda (0)}{\gamma_0 k}\right)^{\frac{1}{k-1}}.
\]
\end{thm}
\begin{pf}
Using the convexity of $  p_l(\beta, \mu) $ and $p^{\mathrm
{appr}}_l(\beta, c_{0}, \mu) $ with respect to $ \lambda (0)$, we get
from a Griffiths theorem ~\cite{grif}
\[
\rho_{0,l}(\mu ) = -\partial_{\lambda (0)} p_l(\beta, \mu) =
-\partial_{\lambda (0)} p^{\mathrm {{appr}}}_l(\beta, c_{0}, \mu) =
\left|c_{0}\right|^2.
\]
Finally, passing to the thermodynamic limit, we obtain for all $\mu>\lambda(0)$,
\[
\lim_{V_l\to \infty}\rho_{0,l}(\mu ) = \rho_{0}(\mu)=
\left|c_{0}\right|^2=\left(\frac{\mu -\lambda (0)}{\gamma_0 k
}\right)^{\frac{1}{k-1}}.
\]
\end{pf}
\qed

6. For the class of Hamiltonians given in equation (\ref{ham1}), in
the framework of the so-called Bogolyubov approximation
~\cite{bog,gin}, it has been given a simple proof of thermodynamic
equivalence of the limit grand-canonical pressures corresponding to
those systems and their respective approximating ones for any integer
$k\ge 2$. Moreover, in contrast to the Bru--Zagrebnov models ~\cite{bru2,bru3}, we prove
that independent on temperature BEC in the sense of macroscopic
occupation of the ground state holds for any integer $k\ge 2$ and any $\mu>\lambda(0)$.
A similar type of BEC is explained entirely by superstability of the model and by absence of an interaction between the ground state occupation number operators and the nonzero modes ones.

\begin{ack}
Partial financial support by Programa MECESUP PUC103 ( Pontificia
Universidad Cat\'olica de Chile), Project ECOS/Conicyt: Estudio
Cualitativo de Sistemas Din\'amicos Cu\'anticos (Chile), Programa de
Mag\'{\i}ster en Matem\'aticas, Universidad de La Serena, Chile and
Program "Fundamental problems of nonlinear dynamics" of the Russian
Academy of Sciences.
\end{ack}

\end{document}